
%
%
\input harvmac

\def\CTPa{\it Center for Theoretical Physics, Department of Physics,
      Texas A\&M University}
\def\CTPb{\it College Station, TX 77843-4242, USA}

\def\HARCa{\it Astroparticle Physics Group,
Houston Advanced Research Center (HARC)}
\def\HARCb{\it The Woodlands, TX 77381, USA}

%
%
%
\def\ie{\hbox{\it i.e.}}     
\def\eg{\hbox{\it e.g.}}

\catcode`\@=11 

\def\lsim{\mathrel{\mathpalette\@versim<}}
\def\gsim{\mathrel{\mathpalette\@versim>}}
\def\@versim#1#2{\vcenter{\offinterlineskip
    \ialign{$\m@th#1\hfil##\hfil$\crcr#2\crcr\sim\crcr } }}
\def\boxit#1{\vbox{\hrule\hbox{\vrule\kern3pt
      \vbox{\kern3pt#1\kern3pt}\kern3pt\vrule}\hrule}}

\def\etal{{\it et. al.}}
\def\r#1{$\bf#1$}

\def\t1{{\tilde 1}}
\def\ov{\overline}

\def\JL{J. L. Lopez}
\def\DVN{D. V. Nanopoulos}

\def\GeV{\,{\rm GeV}}

\def\NPB#1#2#3{Nucl. Phys. B {\bf#1} (19#2) #3}
\def\PLB#1#2#3{Phys. Lett. B {\bf#1} (19#2) #3}

\def\PRL#1#2#3{Phys. Rev. Lett. {\bf#1} (19#2) #3}
\def\PRT#1#2#3{Phys. Rep. {\bf#1} (19#2) #3}

\def\TAMU#1{Texas A \& M University preprint CTP-TAMU-#1}

\nref\Amati{For a review see, \eg, D. Amati, \etal, \PRT{162}{88}{169}.}
\nref\Konishi{K. Konishi, \PLB{135}{84}{439}.}
\nref\Lance{L. Dixon, in Proceedings of The Rice Meeting, ed. by B.
Bonner and H. Miettinen (World Scientific, 1990), p. 811, and references
therein.}
\nref\Dine{I. Affleck, M. Dine, and N. Seiberg, \NPB{256}{85}{557} and
references therein.}
\nref\Ferrara{S. Ferrara, D. L\"ust, A. Shapere, and S. Theisen,
\PLB{225}{89}{363}; S. Ferrara, D. L\"ust, and S. Theisen, \PLB{233}{89}{147}
and \PLB{242}{90}{39}.}
\nref\LN{J. L. Lopez and \DVN, \PLB{251}{90}{73}.}
\nref\guts{See \eg, G. G. Ross, {\it Grand Unified Theories},
(Benjamin/Cummings, MA, 1983); C. Kounnas, A. Masiero, D. V. Nanopoulos and K.
A. Olive, {\it Grand Unification With and Without Supersymmetry and
Cosmological Implications}, (World Scient. Publ. Comp., Singapore, 1984).}
\nref\tc{For a review see E. Farhi and L. Susskind, \PRT{74}{81}{277}.}
\nref\sugra{A. Lahanas and \DVN, \PRT{145}{87}{1}.}
\nref\Witten{J. P. Derendinger, L. Ib\'a\~nez, and H. Nilles,
\PLB{155}{85}{65};
M. Dine, R. Rohm, N. Seiberg, and E. Witten, \PLB{156}{85}{55}.}
\nref\cryp{J. Ellis, \JL, and \DVN, \PLB{247}{90}{257}.}
\nref\KLN{S. Kalara, J. Lopez, and \DVN, \PLB{245}{90}{421},
\NPB{353}{91}{650}.}
\nref\Schwarz{See \eg, J. Schwarz, Caltech preprint CALT-68-1581 (1989).}
\nref\others{
I. Antoniadis, J. Ellis, A. B. Lahanas, and \DVN, \PLB{241}{90}{24};
C. P. Burguess and F. Quevedo, \PRL{64}{90}{2611}; S. Ferrara, N. Magnoli,
T. R. Taylor, and G. Veneziano, \PLB{245}{90}{409}; H. P. Nilles and
 M. Olechowsky, \PLB{248}{90}{268}; P. Binetruy and M. K. Gaillard,
\PLB{253}{91}{119}; M. Cvetic, \etal, \NPB{361}{91}{194}.}
\nref\LT{D. L\"ust and T. Taylor, \PLB{253}{91}{335}.}
\nref\CCM{B. de Carlos, J. A. Casas, and C. Mu\~noz, \PLB{263}{91}{248};
J. A. Casas and C. Mu\~noz, CERN preprint CERN-TH.6187/91.}
\nref\Jan{J. Louis, SLAC preprint SLAC-PUB-5645 (1991).}
\nref\KLNp{S. Kalara, \JL, and \DVN, in preparation.}
\nref\Wetterich{J. Ellis, S. Kalara, K. Olive, and C. Wetterich,
\PLB{228}{89}{264}.}
\nref\FFF{I. Antoniadis, C. Bachas, and C. Kounnas, Nucl. Phys. B
{\bf 289} (1987) 87; I. Antoniadis and C. Bachas, Nucl. Phys. B {\bf298} (1988)
586; H. Kawai, D.C. Lewellen, and S.H.-H. Tye, Phys. Rev. Lett. {\bf57} (1986)
1832; Phys. Rev. D {\bf34} (1986) 3794; Nucl. Phys. B {\bf288} (1987) 1;
R. Bluhm, L. Dolan, and P. Goddard, Nucl. Phys. B {\bf309} (1988) 330;
H. Dreiner, J. L. Lopez, D. V. Nanopoulos, and D. Reiss, Nucl. Phys. B
{\bf 320} (1989) 401.}
\nref\revamp{I. Antoniadis, J. Ellis, J. Hagelin, and \DVN, \PLB{231}{89}{65}.}
\nref\LRT{G. Leontaris, J. Rizos, and K. Tamvakis, \PLB{243}{90}{220}.}
\nref\Lacaze{V. Kaplunovsky, \NPB{307}{88}{145}; I. Antoniadis, J. Ellis,
R. Lacaze, and \DVN, CERN preprint CERN-TH.6136/91 (To appear in Phys. Lett.
B).}
\nref\thresholds{S. Kalara, \JL, and \DVN, \TAMU{46/91} (To appear in Phys.
Lett. B).}
\nref\RT{J. Rizos and K. Tamvakis, \PLB{251}{90}{369}.}
\nref\STAB{J. L. Lopez and \DVN, \PLB{256}{91}{150}.}

\leftline{\titlefont TEXAS A\&M UNIVERSITY}
\leftline{\bf CENTER FOR THEORETICAL PHYSICS}
\Title{\vbox{\baselineskip12pt\hbox{CTP--TAMU--69/91}\hbox{ACT--48}}}
{Gauge and Matter Condensates in Realistic String Models}
\centerline{S. KALARA, JORGE~L.~LOPEZ,\footnote*{Supported
by an ICSC--World Laboratory Scholarship.} and D.~V.~NANOPOULOS}
\bigskip
\centerline{\CTPa}
\centerline{\CTPb}
\centerline{and}
\centerline{\HARCa}
\centerline{\HARCb}
\vskip .3in
\centerline{ABSTRACT}
We examine the inter-relationship of the superpotential containing
hidden and observable matter fields and the ensuing condensates in free
fermionic string models. These gauge and matter condensates of the strongly
interacting hidden gauge groups play a crucial role in the determination of
the physical parameters of the observable sector. Supplementing the above
information with the requirement of modular invariance, we find that a generic
model with only trilinear superpotential allows for a degenerate (and sometimes
pathological) set of vacua. This degeneracy may be lifted by higher order terms
in the superpotential. We also point out some other subtle points that may
arise in calculations of this nature. We exemplify our observations by
computing
explicitly the modular invariant gaugino and matter condensates in the flipped
$SU(5)$ string model with hidden gauge group $SO(10)\times SU(4)$.
\bigskip
\Date{September, 1991}

\newsec{Introduction and General Remarks}
Nonperturbative dynamics of strongly interacting supersymmetric gauge theories
has been explored using a wide variety of tools with great success
\refs{\Amati,\Konishi}. A better understanding of dynamics may lead to
solutions to some deep-rooted problems like supersymmetry breaking \Lance,
the gauge hierarchy problem \Dine, and even string compactification \Ferrara.
In the context of string theory, the task
of incorporating strongly interacting gauge theory dynamics becomes even more
compelling since many crucial properties of the string theory may not be
discerned until a deeper grasp of the vacuum structure of this kind of
gauge theories is obtained \LN.

Typically a theory which aims to go beyond the Standard Model and explain some
of its features, may contain many different mass scales beyond $M_Z$ and may
involve many other gauge degrees of freedom, some of which may even be strongly
interacting. Examples of such theories include all grand unified theories
\guts,
technicolor theories \tc, and supergravity theories with or without strongly
interacting hidden sectors \sugra. In any of these theories one may propose the
existence of a strongly interacting gauge theory for no other reason, save
the problem at hand, \ie, elimination of elementary Higgs fields \tc,
supersymmetry breaking \Witten, etc. However, in string theory one finds that
the consistency of the
theory forces upon us the existence of a completely hidden and/or semi-hidden
gauge group which has to reckoned with. Furthermore, one finds that in a large
class of models, the theory also contains hidden matter fields. The interaction
of the hidden matter fields with themselves and with the observable matter
fields is in principle completely calculable \cryp. It is the interplay between
nonperturbative dynamics of the hidden sector and its possible effect on the
observable sector which is of great interest.

In this paper we examine the reciprocal influence of the nonperturbative
dynamics of strongly interacting gauge theories and the interaction among the
matter fields in a class of models. The type of models we consider are
characterized by two mass scales: $\Lambda_G$ where the gauge theory becomes
strong, and the mass scale ${\cal M}$ (${\cal M}\gg\Lambda_G$). The key
ingredient that
allows us to probe the theory in great detail is the tamed ultraviolet behavior
of the theory due to supersymmetry and the presence of genuinely stringy
discrete symmetries.

In a typical supergravity theory, without the benefit of any string input,
the dynamics of the gaugino condensate depends primarily on the gauge coupling
constant $g$ and the masses of the matter fields charged under the strongly
interacting gauge group. However, in string theory the gauge coupling $g$ is
related to the vacuum expectation value of the dilaton field $S$ as
$1/g^2=4{\rm Re}\vev{S}$ and consequently becomes a dynamical variable.
Furthermore, the interactions of the matter fields are only dictated by the
string dynamics and, in a class of theories, are calculable \KLN. Additionally,
string theory imposes very strong discrete symmetries (the so-called target
space modular symmetries \Schwarz) on the theory, making the structure of the
gaugino condensate very tightly constrained.

In the preparatory examples treated in the literature
\refs{\others,\LT,\CCM,\Jan},
\ie, examples with
only one modulus field and the case where all matter fields of the hidden
gauge group acquire masses through trilinear superpotential terms, most of
the stringy information has been incorporated. Specifically, for the case of
$SU(N)$ with $M$ flavors in the fundamental representation, when the mass
${\cal M}$ of the matter fields is ${\cal M}\ll\Lambda_G$, one finds \LT
\eqna\X
$$\eqalignno{{Y^3\over32\pi^2}&=(32\pi^2e)^{M/N-1}[c\eta(T)]^{2M/N-6}
[{\rm det}\,{\cal M}]^{1/N} e^{-32\pi^2S/N},&\X a\cr
\Pi_{ij}&={Y^3\over32\pi^2}{\cal M}^{-1}_{ij},&\X b\cr}$$
where $T$ is the modulus field and $\eta$ the Dedekind function, $c$ is an
unknown constant, the composite superfield $Y$ is defined as
$Y^3=W_\alpha W^\alpha/S^3_0$ where $S_0$ is the chiral compensator,
and $\Pi_{ij}=\vev{C_iC_j}$ are the matter condensates.
Using one-loop renormalization group equations, the complementary case of
${\cal M}\gg\Lambda_G$ can also be incorporated \CCM. For an alternate
approach using an effective Lagrangian see Ref. \Jan.
However, a generic full-fledged string theory example introduces its own set
of complications. It is the purpose of this paper to examine some of these
involutions.

In a realistic example, to obtain a coherent picture of the vacuum structure
of the theory requires a multipronged approach. Generically one finds that
some of the matter fields which carry nontrivial gauge quantum numbers of the
strongly interacting theory do not acquire mass at the trilinear level of the
superpotential. As it can be easily shown that a strongly interacting
supersymmetric gauge theory with massless matter fields is beset with the
problem of pathological vacuum structure \Amati, this necessitates that the
mass calculations for the matter fields be carried up to quartic order and
beyond.

The presence of the modular invariance further restricts the type of terms
that can arise \Ferrara. In a class of models higher order nonrenormalizable
terms can be calculated \KLN\ and their modular invariance properties inferred
\KLNp. The
key point to note here is that the requirement of a stable vacuum mandates that
the superpotential be probed up to the level at which the determinant of the
mass matrix is nonzero. The presence of a zero eigenvalue in the mass matrix
invariably leads to both theoretical (unstable vacua \Amati) and experimental
(breakdown of the equivalence principle \Wetterich) difficulties.

Furthermore, strongly interacting supersymmetric gauge theories are notorious
for containing a large class of degenerate vacua. Unless the effects of the
superpotential are taken into account, the degeneracy remains unbroken.
Additionally, many of the expositions in strongly interacting gauge theory
dynamics are based on the global and/or local symmetries of the theory \Amati,
thus in the absence of complete knowledge of the superpotential an unambiguous
identification of the vacuum structure may not be possible.

The free-fermionic formulation of string theory \FFF\ is specially suited to
explore these questions, since the nonrenormalizable terms can be explicitly
calculated \KLN\ and modular properties of these terms can also be determined.
Succintly put,
a nonrenormalizable term $\Phi_1\ldots\Phi_N$ can be calculated by evaluating
the correlator $\vev{\Phi_1\ldots\Phi_N}$. After taking into account the
massless exchanges, a nonzero piece at low momenta signifies a presence of
such a term in the superpotential. A priori, the existence of such a term
in the superpotential would be inconsistent with modular invariance. However,
it can be shown \KLNp\ that the nonzero value of the correlator
$\vev{\Phi_1\ldots\Phi_N}$ implies a nonzero value of the following series of
correlators $\vev{T^p\Phi_1\ldots\Phi_N}$, where $p\ge N-3$ and $T^p$ is a
definite product of $p$ moduli fields of the theory. Supplementing this
observation with the requirement of modular invariance, we find that the
coefficient of a given nonrenormalizable term can be summed up into a
modular covariant function made up of products of Dedekind $\eta$ functions of
the moduli fields involved. A modular invariant gaugino condensate will
also have to be reconciled with the nontrivial modular properties of
nonrenormalizable terms.

Thus we see that the question of gaugino condensate cannot be answered
independently of the superpotential. The effect of the superpotential is
felt in determining the vacuum structure of the theory lifting the
degeneracy of the vacuum, and in the specific value of the gaugino
condensate.
\newsec{Application to Flipped $SU(5)$}
We now present an explicit example of the calculation of hidden massive
matter condensates and the various subtle points that arise in realistic string
models. To this end we study the flipped $SU(5)$ string model \revamp\ which
has
a rich hidden sector spectrum composed of two gauge groups, namely
$SO(10)$ and $SU(4)$. Since $SO(10)$ is expected to become strongly
interacting at a scale ($\Lambda_{10}\approx10^{14-16}\GeV$ \LRT) much higher
than the respective $SU(4)$ scale ($\Lambda_4\approx10^{10-12}\GeV$ \LRT),
in the following we deal exclusively with $SO(10)$ condensates. Besides
$\Lambda_{10}$, two other mass scales  come into play: the $SU(5)\times U(1)$
beaking scale, $V,\ov V\sim10^{15-16}\GeV$; and the scale of singlet vevs
needed to cancel the anomalous D-term, $\vev{\phi}\sim10^{17}\GeV$
\refs{\revamp,\LN}. All these scales are to be compared with the string
unification scale $M_{SU}\approx1.24\times g\times10^{18}\GeV$
\refs{\Lacaze,\thresholds}\ and
the scale of nonrenormalizable terms in the superpotential
$M\approx10^{18}\GeV$ \KLN.

We first need to determine the mass matrix for the $SO(10)$ fields. These
belong to the \r{10} of $SO(10)$ and are denoted by $T_i,\,i=1\to5$. There
are generally three sources of mass terms for these fields:
\medskip
\noindent(i) $T_iT_j\phi^{N-2}$: at N=3 we have the following terms
\refs{\LN,\RT}
\eqn\XX{\ov\Phi_{23}T^2_1,\Phi_{31}T^2_2,\Phi_{23}T^2_4,\Phi_{31}T^2_5,
                                                                \phi_3 T_4T_5,}
where the various $\phi$'s are singlet fields which will generally
get vevs $\vev{\phi}\sim10^{17}\GeV$, but could vanish for consistency or
phenomenological reasons. Following the methods in \STAB, it can be shown
that no new terms arise at any order $N\ge4$, except for small corrections
to the above terms (\eg, $T^2_1\Phi_{31}\bar\phi^2_1/M^2$).
\smallskip
\noindent(ii) $T_iT_j F_{1,3}\bar F_5\phi^{N-4}$: here $F_{1,3}$ and
$\bar F_5$ are the $SU(5)$ fields which get vevs $V,\ov V$ respectively.
No such terms occur at N=4,6,8. At N=5 we get $T_1T_4F_1\bar F_5\phi_3/M^2$
and at N=7
\eqna\XXi
$$\eqalignno{&T_3T_4F_3\bar F_5\{\phi_2\phi_3\bar\phi^-,\phi_2\bar\phi_3\phi^+,
                \phi_3\bar\phi_2\phi^+\}/M^4,&\XXi a\cr
&T_3T_5F_3\bar F_5\{\phi_3\bar\phi^-\Phi_{31},\bar\phi_3\phi^+\Phi_{31}\}/M^4.
                                &\XXi b\cr}$$
Yet higher order terms ($N\ge9$) are further suppressed by powers of
$(\vev{\phi}/M)\sim10^{-2}$.
\smallskip
\noindent(iii) $T_iT_jT_kT_l\phi^{N-4}$: no such terms occur at N=4,5,7.
At N=6 we get
\eqna\XXii
$$\eqalignno{&T_3T_3T_4T_4\phi_{45}\phi^+/M^3,&\XXii a\cr
        &T_1T_1T_4T_5\{\phi_1\bar\phi_2,\phi_2\bar\phi_1\}/M^3,&\XXii b\cr
        &T_1T_1T_4T_5\phi_1\Phi_{31}/M^3.&\XXii c\cr}$$
With yet higher orders further suppressed.

We should point out that a very useful constraint in searching for
non-vanishing higher-order nonrenormalizable terms in the superpotential
is given by the modular invariance of the all-orders superpotential, as
discussed above.

The resulting $T_i$ mass matrix can be written as follows
\eqn\XXiii{
{\cal M}_{ij}=\pmatrix{\bar\Phi_{23}&0&0&\eta&0\cr
        0&\Phi_{31}&0&0&0\cr
        0&0&\delta_{44}&\delta_{34}+\epsilon_4&\epsilon_5\cr
        \eta&0&\delta_{34}+\epsilon_4&\delta_{33}+\Phi_{23}&\phi_2\cr
        0&0&\epsilon_5&\phi_2&\Phi_{31}\cr}}
where
\eqna\XXiv
$$\eqalignno{\delta_{ij}&=\vev{T_iT_j}\phi_{45}\phi^+{1\over M^3}
        \equiv {\delta\over M}\Pi_{ij},&\XXiv a\cr
        \epsilon_{4,5}&=F_3\bar F_5{\phi^3_{4,5}\over M^4},&\XXiv b\cr
        \eta&=F_1\bar F_5\phi_3{1\over M^2},&\XXiv c\cr}$$
with the relevant $\phi^3_{4,5}$ given in \XXi{}. It is important to estimate
the sizes of the various entries in ${\cal M}_{ij}$. With the above mentioned
scales we obtain: $\vev{\phi}\sim10^{-1}M$, $\eta\sim10^{-5}M$,
$\epsilon_{4,5}\sim10^{-7}M$, and $\delta\sim10^{-2}$.

Since $\Pi_{ij}={Y^3\over32\pi^2}{\cal M}^{-1}_{ij}$, and ${\cal M}_{ij}$
depends on $\Pi_{ij}$ also, the solution of the resulting equations can be
rather nontrivial. We now invert the matrix ${\cal M}$ in various levels
of approximation to exhibit the subtleties that can arise. As a first
approximation let us drop all nonrenormalizable contributions to ${\cal M}$.
In this limit the matrix breaks up into three blocks with a zero eigenvalue
for $T_3$. The solution is
$\Pi_{11,22,44,45,55}\sim{Y^3\over32\pi^2}{1\over\vev{\phi}}$ and
$\Pi_{33}\to\infty$, with all the other condensates vanishing. Clearly this
is an unphysical situation and we are forced to consider nonrenormalizable
terms.

Next we keep the $\delta_{ij}$ terms and neglect the $\epsilon_{4,5}$ and
$\eta$ terms. Schematically the matrix becomes
\eqn\XXv{\pmatrix{\phi&0&0&0&0\cr
                0&\phi&0&0&0\cr
                0&0&\delta_{44}&\delta_{34}&0\cr
                0&0&\delta_{34}&\delta_{33}+\phi&\phi_2\cr
                0&0&0&\phi_2&\phi\cr}.}
The $2\times2$ submatrix involving $T_{1,2}$ remains unchanged, \ie,
$\Pi_{11}\sim\Pi_{22}\sim{Y^3\over32\pi^2}{1\over\vev{\phi}}$. To simplify
the calculation without loss of generality, let us set $\vev{\phi_2}=0$,
thus decoupling the $T_5$ field, \ie, $\Pi_{45}=0$,
$\Pi_{55}\sim{Y^3\over32\pi^2}{1\over\vev{\phi}}$. The remaining $2\times2$
matrix can be easily inverted and the following three equations result
\eqna\XXvi
$$\eqalignno{\Pi_{44}\left\{1-{Y^3\over32\pi^2}{1\over D}{\delta\over M}
                                \right\}&=0,&\XXvi a\cr
 \Pi_{33}\left\{1-{Y^3\over32\pi^2}{1\over D}{\delta\over M}
                \right\}&={Y^3\over32\pi^2}{\vev{\phi}\over D},&\XXvi b\cr
 \Pi_{34}\left\{1+{Y^3\over32\pi^2}{1\over D}{\delta\over M}
                                \right\}&=0,&\XXvi c\cr}$$
where
\eqn\XXvii{D=\left(\vev{\phi}+{\delta\over M}\Pi_{33}\right){\delta\over M}
        \Pi_{44}-\left({\delta\over M}\Pi_{34}\right)^2.}
Equation \XXvi{c} can be solved if (i) $\Pi_{34}=0$ and/or (ii)
${Y^3\over32\pi^2}{\delta\over DM}=-1$. In case (ii) we get from \XXvi{a}
\eqn\XXX{\Pi_{44}=0,}
and from \XXvi{b}
\eqn\XXXi{\Pi_{33}={1\over2}{Y^3\over32\pi^2}{\vev{\phi}\over D}=
-{1\over2}{M\over\delta}\vev{\phi}\sim M^2.}
Also, $\Pi_{34}$ gets determined through $D$,
\eqn\XXXii{\Pi_{34}=\left({Y^3\over32\pi^2}{M\over\delta}\right)^{1/2}.}
In case (i), Eqn. \XXvi{a}\ gives $\Pi_{44}=0$ and/or
${Y^3\over32\pi^2}{\delta\over DM}=1$, both of which lead to unphysical or
inconsistent solutions. Thus solution (ii) is preferred. Note that the field
$T_3$ remains very light ($\delta_{44}=0$), while $\Pi_{33}\sim M^2$ is
finite.

For the case when all matter fields are the mass eigenstates, the
usual clustering argument, which is based on the symmetries of the
superpotential, would imply that $\Pi_{34}=0$. However, we see that $T_3,T_4$
are not mass eigenstates due to a mixing term
$\epsilon_4T_3T_4\sim10^{-7}MT_3T_4$. The presence of such a term breaks the
degeneracy of the vacuum: $\Pi_{44}=0,\Pi_{34}\not=0$ and
$\Pi_{44}\not=0,\Pi_{34}=0$. One can repeat the above calculations taking
$\epsilon_4\not=0$ (and $\epsilon_5=0$ for simplicity) and the only change
in Eqs. \XXvi{}\ is in the right-hand side of Eq. \XXvi{c}\ which becomes
$-{Y^3\over32\pi^2}{\epsilon_4\over D}$. The solutions of the equations are
more complicated, but they differ negligibly from the case $\epsilon_4=0$.
That is, even though $\epsilon_4$ is negligible numerically, its presence
determines the correct vacuum of the theory.

The above example shows the importance of knowing the higher-order
superpotential
to obtain physically acceptable results. We should point out that it may
happen that some of the singlet vevs vanish, implying $\delta=0$ and/or
$\epsilon_{4,5}=0$. As discussed above, this scenario may lead to unphysical
solutions and thus these vev choices appear disfavored.
\newsec{Calculation of the Gaugino Condensate}
We now calculate the only unknown left in the theory, namely the gaugino
condensate $Y^3$. Equation \X{a}\ has been obtained by imposing the modular
symmetry of $SL(2,{\bf Z})$ on the effective action. Generally the modular
group is much larger. For the class of free-fermionic models under
consideration
one finds that the symmetry group is $SL(2,{\bf Z})^3$ accompanied by three
moduli fields \KLNp. Correspondingly
the gaugino condensate is straightforwardly generalized, and in the case of
$SO(10)$ with $M$ massive flavors we obtain
\eqn\XXviii{{Y^3\over32\pi^2}=(32\pi^2e)^{M/8-1}
[c\eta(T_1)\eta(T_2)\eta(T_3)]^{-2}[{\rm det}\,{\cal\widehat M}]^{1/8}
e^{-32\pi^2S/8},}
where ${\rm det}\,{\cal\widehat M}$ is the determinant of the mass matrix
${\cal M}$ calculated with modular invariant mass terms. This way $Y^3$ has
modular weight $(-1,-1,-1)$ and $W_{eff}\propto Y^3$ \Jan\ has the correct
modular properties in this generalized case.

Let us now compute ${\rm det}\,{\cal\widehat M}$ for the case analyzed in
Eqs. \XXX--\XXXii\ above where
\eqn\L{{\rm det}\,{\cal M}=\ov\Phi_{23}\Phi_{31}^2
[\delta_{44}\delta_{33}-\delta^2_{34}+\delta_{44}\Phi_{23}].}
The trilinear masses $\ov\Phi_{23}$ and $\Phi_{31}$ have modular
weights\foot{The modular weights of fields in free fermionic models are
given in \thresholds. Here we also use $T_3,\phi_{45},\phi^+:\,(-1/2,-1/2)$
and $T_4:\,(0,-1/2)$. Note that we only consider the modular properties under
the $T_{1,2}$ moduli fields. The modular transformations under $T_3$ are less
obvious and are still
under investigation.} $(-1,0)$ and $(0,-1)$ respectively and thus we take
instead $\eta^2(T_1)\ov\Phi_{23}$ and $\eta^2(T_2)\Phi_{31}$. The $\delta_{ij}$
terms come from the following superpotential term
\eqn\Li{a\,T_3T_3T_4T_4\phi_{45}\phi^+\,\eta^2(T_1)\eta^4(T_2),}
where $a$ is an ${\cal O}(1)$ calculable constant and
the $\eta$ functions insure that this contribution to the superpotential
has the appropriate modular weight (\ie, $(-1,-1)$) \KLNp. This implies
that $\delta/M$ defined in Eq. \XXiv{a}\ should be multiplied by $\eta^2(T_1)
\eta^4(T_2)$ as well. With this information and the modular weights of the
fields involved, one can readily determine whether additional $\eta$ functions
are needed to make the mass terms in \L\ modular invariant. The final result is
\eqna\Lii
$$\eqalignno{{\rm det}\,{\cal\widehat M}&=\ov\Phi_{23}\Phi^2_{31}\eta^2(T_1)
\eta^4(T_2)
[(\delta_{33}\delta_{44}-\delta_{34}^2)\eta^2(T_1)-\delta_{44}\Phi_{23}],
                                                        &\Lii a\cr
        &=-\ov\Phi_{23}\Phi^2_{31}\,a\,{\phi_{45}\phi^+\over M^3}
                {Y^3\over32\pi^2}\eta^6(T_1)\eta^8(T_2),&\Lii b\cr}$$
where Eqs. \XXX\ and \XXXii\ have been used.

To make our results for ${Y^3\over32\pi^2}$ more transparent, we insert the
missing units in Eq. \XXviii\ and set $M=5$. The exponential factor then
becomes $M^{19/8}e^{-32\pi^2S/8}$ with $M\approx10^{18}\GeV$. Using $4g^2=1/S$
and introducing the $SO(10)$ condensation scale
$\Lambda_{10}=Me^{8\pi^2/(\beta g^2)}$, with $\beta=-3\times8+5=-19$, this
factor becomes $\Lambda_{10}^{19/8}$, and thus we can write
\eqn\Liii{{Y^3\over32\pi^2}=(32\pi^2e)^{-3/8}[c\eta(T_1)\eta(T_2)]^{-2}
\Lambda_{10}^3\left({\rm det}{{\cal\widehat M}\over\Lambda_{10}}\right)^{1/8},}
which is a modular invariant generalization of the usual gaugino condensate
expression \Amati. Note that since ${\rm det}{\cal\widehat M}$ depends on
${Y^3\over32\pi^2}$ as well, the final expression for ${Y^3\over32\pi^2}$
is different, as follows
\eqn\Liv{{Y^3\over32\pi^2}=(32\pi^2e)^{-3/7}[c\eta(T_1)\eta(T_2)]^{-2}
\Lambda^3_{10}\left({\rm det}{{\cal\widehat
M}'\over\Lambda_{10}}\right)^{1/7},}
with
\eqn\Lv{{\rm det}{{\cal\widehat M}'\over\Lambda_{10}}=-\ov\Phi_{23}\Phi^2_{31}
a{\phi_{45}\phi^+\over\Lambda^2_{10}M^3}\,\eta^4(T_1)\eta^6(T_2).}
Numerically, the
$\Lambda^3_{10}({\rm det}{{\cal\widehat M}'\over\Lambda_{10}})^{1/7}$ terms
determine the scale of ${Y^3\over32\pi^2}$. For typical values of the
parameters we obtain $\Lambda_{10}\sim10^{15}\GeV$,
${\rm det}{{\cal\widehat M}'\over\Lambda_{10}}\sim10$, and
${Y^3\over32\pi^2}\sim(10^{15}\GeV)^3$. The matter condensates then become
$\Pi_{11}\sim\Pi_{22}\sim\Pi_{55}\sim(10^{14}\GeV)^2$, $\Pi_{44}\sim0$,
$\Pi_{33}\sim(10^{18}\GeV)^2$, and $\Pi_{34}\sim(10^{16}\GeV)^2$.
\newsec{Conclusions}
Gauge and matter condensates are very important in the analysis of realistic
string-derived models, since these typically involve hidden gauge groups which
communicate with the observable sector, although rather feebly. The
calculations presented in this paper address some of the difficulties that
are likely to arise in typical models, such as the degeneracy of the vacuum
and the need to explore the superpotential to high orders. The latter requires
an understanding of the modular invariant properties of the superpotential to
all orders in nonrenormalizable terms.
We have made the above points explicit in the particular case of the flipped
$SU(5)$ model with $SO(10)$ hidden gauge group and matter fields in the
fundamental representation. In this case the gaugino condensate has to be
solved self-consistently together with the matter condensates, a feature that
may only be appreciated in a calculation in an explicit string model.
\bigskip
\noindent{\it Acknowledgments}: This work has been supported in
part by DOE grant DE-FG05-91-ER-40633.
\listrefs
\bye